# The emergence and diversification of dog morphology

Allowen Evin[1,2]*†, Carly Ameen[2]*†, Colline Brassard[3,4], Sophie Dennis[5], Ekaterina E. Antipina[6], Vincent Bonhomme[1,7], Myriam Boudadi-Maligne[8], Kate Britton[9], Francisco Gil Cano[10], Ruth F. Carden[11,12], Julien Claude[1], Lídia Colominas[13], Stefan Curth[14], Sergey Egorovich Fedorov[15], Joan Frances[16], Daniela C. Kalthoff[17], Andrew C. Kitchener[18,19], Rick Knecht[9], Pavel Kosintsev[20,21], Anna Linderholm[22,23], Robert Losey[24], Ilia Merts[25], Viktor Merts[25], Maria Mostadius[26], Mark Omura[27], Vedat Onar[28], Alan K. Outram[2], Joris Peters[29,30], André Rehazek[31], Erika Rosengren[32], Mikhail Sablin[33], Paul Sciulli[34], Maria Seguí[35], Z. Jack Tseng[36,37], Emma Usmanova[38,39], Victor Varfolomeev[40], Susan Crockford[41], Yaroslav Kuzmin[42], Laurent Frantz[5,43], Keith Dobney[9,44], Greger Larson[23]

[1]ISEM, University of Montpellier, CNRS, EPHE, IRD, Montpellier, France. [2]Department of Archaeology and History, University of Exeter, Exeter, UK. [3]UMR 7179 MECADEV (Mécanismes Adaptatifs et Evolution), Muséum National d'Histoire Naturelle, Paris, France. [4]VetAgro Sup Lyon, Marcy l'Etoile, France. [5]School of Biological and Behavioural Sciences, Queen Mary University of London, London, UK. [6]Institute of Archaeology, Russian Academy of Sciences, Moscow, Russia. [7]Athéna, Roquedur, France. [8]CNRS, Bordeaux University, UMR5199 PACEA, Bordeaux, France. [9]Department of Archaeology, University of Aberdeen, Aberdeen, Scotland, UK. [10]Departmento de Anatomía y Anatomía Patológica Comparadas, Facultad de Veterinaria, Universidad de Murcia, Murcia, Spain. [11]School of Archaeology, University College Dublin, Dublin, Ireland. [12]Danu Ruadh Teoranta, Cavan, Ireland. [13]Institut Català d'Arqueologia Clàssica, Tarragona, Spain. [14]Aquazoo Löbbecke Museum, Düsseldorf, Germany. [15]Mammoth Museum of North-Eastern Federal University, Yakutsk, Russia. [16]Universitat Autònoma de Barcelona, Barcelona, Spain. [17]Department of Zoology, Swedish Museum of Natural History, Stockholm, Sweden. [18]Department of Natural Sciences, National Museums Scotland, Edinburgh, Scotland, UK. [19]School of Geosciences, University of Edinburgh, Edinburgh, Scotland, UK.

[20]Ural Federal University, Ekaterinburg, Sverdlovsk Oblast, Russia. [21]Institute of Plant and Animal Ecology, Russian Academy of Sciences, Ekaterinburg, Russia. [22]Centre for Palaeogenetics, Department of Geological Sciences, Stockholm University, Stockholm, Sweden. [23]Palaeogenomics and Bio-Archaeology Research Network, School of Archaeology, University of Oxford, Oxford, UK. [24]Department of Anthropology, University of Alberta, Edmonton, AB, Canada. [25]Toraighyrov University, Joint Research Center for Archeological Studies, Pavlodar, Kazakhstan. [26]The Biological Museum, Lund University, Lund, Sweden. [27]Museum of Comparative Zoology, Harvard University, Cambridge, MA, USA. [28]Muğla Sıtkı Koçman University, Milas Faculty of Veterinary Medicine, Milas, Muğla, Turkey. [29]Bavarian Natural History Collections, State Collection of Palaeoanatomy Munich (SPM), Munich, Germany. [30]Institute of Palaeoanatomy, Domestication Research, and the History of Veterinary Medicine, Ludwig Maximilian University Munich, Munich, Germany. [31]Naturhistorisches Museum Bern, Bern, Switzerland. [32]Historical Museum, Lund University, Lund, Sweden. [33]Zoological Institute of the Russian Academy of Sciences, St. Petersburg, Russian Federation. [34]Department of Anthropology, Ohio State University, Columbus, OH, USA. [35]Departament de Prehistòria, Universitat Autònoma de Barcelona, Barcelona, Spain. [36]University of California, Berkeley, Berkeley, CA, USA. [37]American Museum of Natural History, New York, NY, USA. [38]Saryarka Archaeological Institute of Karaganda Buketov University, Kazakhstan, Margulan Archaeological Institute, Almaty, Kazakhstan. [39]Khalikov Archaeological Institute, Kazan, Russia. [40]Karaganda Buketov University, Karaganda, Kazakhstan.

[41]Pacific Identifications Inc., Victoria, BC, Canada. [42]Sobolev Institute of Geology and Mineralogy, Siberian Branch of the Russian Academy of Sciences (IGM SB RAS), Novosibirsk, Russia. [43]Palaeogenomics Group, Institute of Palaeoanatomy, Domestication Research and the History of Veterinary Medicine, LMU Munich, Munich, Germany. [44]Department of Archaeology, Classics and Egyptology, University of Liverpool, Liverpool, UK.

*Corresponding author. Email: allowen.evin@umontpellier.fr (A.E.); C.Ameen@exeter.ac.uk (C.A.) †These authors contributed equally to this work.

**Abstract:** Dogs exhibit an exceptional range of morphological diversity as a result of their long-term association with humans. Attempts to identify when dog morphological variation began to expand have been constrained by the limited number of Pleistocene specimens, the fragmentary nature of remains, and difficulties in distinguishing early dogs from wolves on the basis of skeletal morphology. In this study, we used three-dimensional geometric morphometrics to analyze the size and shape of 643 canid crania spanning the past 50,000 years. our analyses show that a distinctive dog morphology first appeared at about 11,000 calibrated years before present, and substantial phenotypic diversity already existed in early Holocene dogs. Thus, this variation emerged many millennia before the intense human-mediated selection shaping modern dog breeds beginning in the 19th century.

The origins and development of domestic dogs remains one of the most debated topics in archaeology. Multiple lines of evidence suggest that domestic dogs emerged during the Late Pleistocene, and palaeogenomic data show that major dog lineages had already diverged before 11,000 years before present (BP) (*1–3*). The burial of a young, diseased puppy and an adult canid alongside a human from the northern European site of Bonn-Oberkassel ~15,000 years ago has been interpreted as evidence for the canids' domestic status (*4*). The earliest direct genomic evidence derived from a domestic dog is based on a nuclear genome recovered from a 10,900 year BP Mesolithic specimen from Veretye in northwest Russia (*3*, *5*).

Several Late Pleistocene specimens dated to as early as 35,500 years BP from across Eurasia were tentatively identified as dogs based on their distinct morphologies (*6*), although to date, all of those assessed have had wolf genomic signatures (*7*, *8*). Recent biometric studies have further challenged the domestic status of these ancient individuals previously identified as dogs, and some exhibit wolf-like skulls (*9–11*). Studies of more recent archaeological crania have revealed significant morphological diversity in ancient dog populations, and several traits have no modern equivalent (*12*). These results highlight the need for a large-scale investigation of both wild and domestic canid variation since the Late Pleistocene that is not limited by geographical or temporal constraints and does not rely exclusively on modern comparative datasets.

Here, to track the emergence and diversification of domestic dog morphologies through time, we applied a geometric morphometric approach capable of identifying and quantifying fine-scale shape changes (*13*). In total, we analyzed a global dataset of 643 modern and archaeological canid crania (Fig. 1A) spanning the past 50,000 years using three-dimensional (3D) digital models and a landmark-based approach (fig. S1, table S1, and dataset S1). Modern wolves (*Canis lupus*; $n = 86$), representing the only available secure wild morphology, originated from Asia ($n = 22$), Europe ($n = 22$), North America ($n = 25$), and Southwest Asia ($n = 10$) (origin was not determined for $n = 2$).

While acknowledging that Pleistocene wolves may exhibit distinct morphologies, several lines of evidence validate the use of modern wolves as reasonable proxies. Genomic data show that they retain 10 to 40% Pleistocene ancestry (*7*), and previous work has reported Pleistocene crania sharing morphometric characteristics with modern wolves [e.g., (*11*)]. In the absence of adequate fossil samples, modern wolves provide the only consistent comparative framework for assessing differences from early dogs.

The modern dog sample ($n = 158$) includes 100 specimens of 47 recognized breeds, 36 street dogs, and 22 individuals of mixed or unknown breed affiliation. Although this sample captures a good proportion of the full diversity of contemporary domestic morphologies, it is not intended to serve as a proxy for ancient domestic forms.

We used boxplots of the log centroid size (Fig. 1B) and a principal components analysis (PCA) for shape (Fig. 1C) to characterize the morphometric variation within four distinct categories: (i) Late Pleistocene specimens dated from 50,000 to 12,700 years BP ($n = 17$), (ii) Holocene specimens from 11,700 calibrated years BP (cal BP) recovered from archaeological sites ($n = 374$), (iii) modern wolves ($n = 86$), and (iv) modern dogs ($n = 158$) (table S1).

On average, both modern wolves and Pleistocene specimens had larger skulls than those of modern dogs and Holocene specimens, although substantial overlap exists among all groups (Fig. 1B). Modern dogs exhibit the greatest size range, slightly exceeding that of Holocene canids, and these groups have a greater variety of small-skulled canids than modern wolves and Pleistocene canids (table S2). Although the Holocene specimens are on average smaller than the other three groups, their overall size range fully overlaps with modern dogs (Fig. 1B).

With respect to cranial shape, Late Pleistocene specimens are distinct from modern wolves

(fig. S2) but show a close overall proximity, as indicated by an overlap on the first two PCA axes (Fig. 1C and table S2). By comparison, mod- ern dogs were found to have statistically distinct and more diverse cranial shapes than those of modern wolves and Pleistocene specimens, a result that corroborates previous studies (*14, 15*) (Fig. 1, B and C, and table S2). The skull shape of Holocene specimens largely overlaps with that of modern dogs, although they also have some unique morphologies that are not represented within our modern dog reference dataset (Fig. 1C). Both modern dog and Holocene canid skull diversity exceed the shape variability found in modern wolves and Pleistocene canids (Fig. 1, B and C, and table S2).

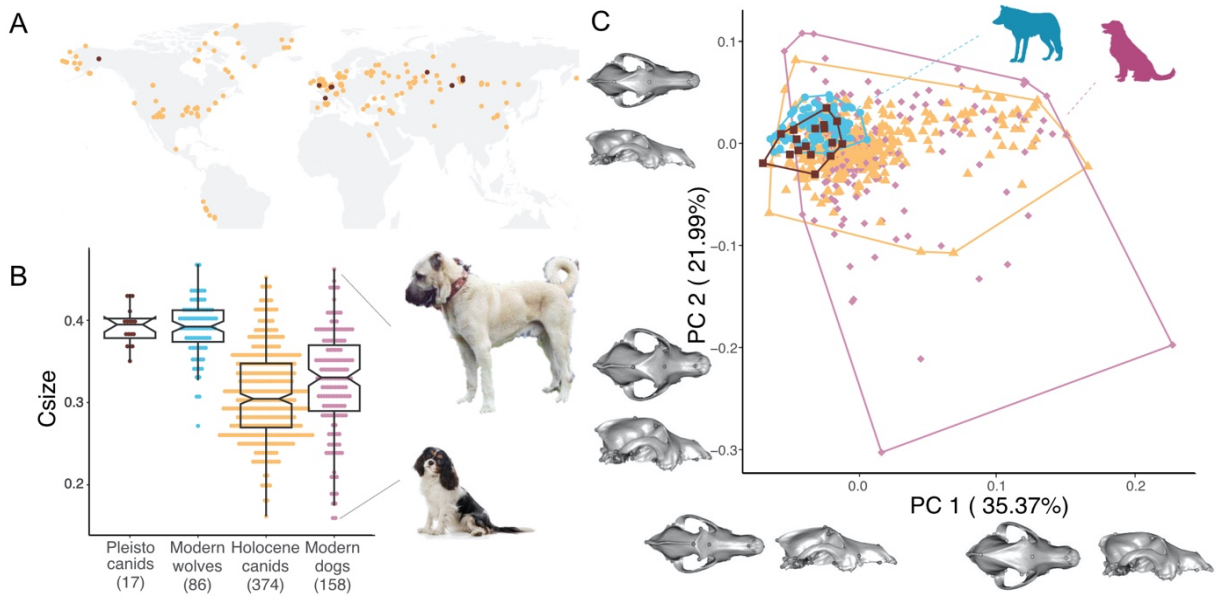

***Fig. 1. Geographic and temporal diversity, skull size, and shape variation of the canid specimens analyzed.*** *(A) Map showing the location of the samples included in this study. Holocene samples are depicted in yellow and Pleistocene canids in brown. (**B**) Boxplot showing skull centroid size variation among Pleistocene canids, modern wolves, Holocene canids, and modern dogs. The extremes of the full modern dog size range are represented by the Aksaray Malaklisi (largest) and the King Charles spaniel (smallest). (**C**) PCA depicting skull shape variation along the two first principal component axes showing the position of the same groups as in (**B**) accompanied by visualizations of the shape changes at the extremes of PC1 and PC2. The colored points in (**B**) and (**C**) represent Pleistocene canids (brown squares), modern wolves (blue dots), Holocene canids (orange triangles), and modern dogs (purple diamonds).*

To test for the presence of domestic canids in the Late Pleistocene, we devel- oped a two-step approach to trace the appearance of domestic skull morphologies in our ancient dataset. First, we calculated the mean skull shape of modern wolves (the only securely identified wild reference specimens) and computed the Procrustes distance of each specimen in our dataset to that average morphology (Fig. 2) (see the materials and methods).

In total, 151 individuals ($n = 81$ Holocene specimens and $n = 70$ modern dogs) had a skull shape that exceeded the distance from the average wolf of all modern wolves and Pleistocene canids (e.g., above the brown line in Fig. 2; also see table S1 and fig. S3). This group, which we refer to as "morphological dogs," can be confidently distinguished from modern wolves with a correct leave-one-out cross-validation percentage of 92.4% [confidence interval (CI): 90.1 to 94.8%; see the materials and methods] (*16*).

Next, we identified the remaining 406 ancient and modern specimens as either wild or domestic by using a predictive discriminant analysis procedure (see the materials and methods), with modern wolves ($n = 86$) and morphological dogs ($n = 151$) as the reference for wild and domestic morphology, respectively (*17*). This approach accounts for the challenges of quantifying ancient variability that may not be fully rep- resented in modern dog morphology, especially for individuals dated to the early phases of the domestication process. This analysis resulted in the assignment of 281 Holocene specimens as morphological dogs and 101 individuals as morphological wolves (table S1). On average, the in- dividuals assigned as morphological dogs had smaller skulls than those of modern wolves, Late Pleistocene canids, and those archaeological specimens identified as morphological wolves (Fig. 3A and fig. S4). This result is consistent with the reduction in size commonly observed in many large mammal species during domestication (*18*).

Of the 17 Late Pleistocene canids, including specimens previously proposed to be early dogs [e.g., those found in Goyet, Belgium (*19, 20*)], were all classified as morphological wolves (Fig. 3A and table S1). The earliest specimens (one directly dated to 11,145 to 10,724 cal BP) catego- rized as morphological dogs in this analysis were three skulls from the Russian Mesolithic site of Veretye (Fig. 3A) (*21*). These specimens are also (thus far) the earliest dogs identified with nuclear genomic data (*3, 5*). Two other Mesolithic specimens from Danish sites (Brabrand Sø and Gudso Vig) dating to the Ertebølle period (7350 to 5900 years BP) were also morphologically classified as dogs (table S1) (*22*).

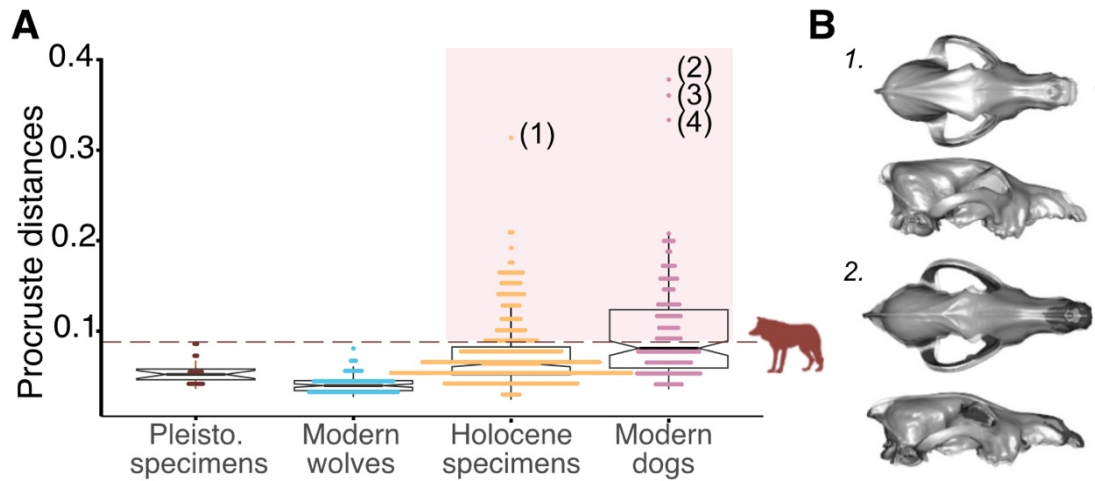

*Fig. 2. Identification of specimens outside the diversity of the wolf.* (A) Procrustes distance between each specimen and the average modern wolf. The maximum distance observed in both modern wolves and Pleistocene canids (brown line) is indicated. Extreme specimens are highlighted as: Holocene canid from Zaojiaoshu (China, 3850 to 3450 BP) (*33*) (1) and specimens of the modern breeds King Charles spaniel (2), Carlin dog (3), and French bulldog (4). Shaded red box indicates specimens that position above the brown line and are classified as morphological dogs. (B) Visualization of the skull shape differences between morphological dogs (1) and modern wolves (2).

In Asia, the earliest morphological dogs in our dataset originate from the Russian Neolithic sites of Shamanka II (7420 to 7325 cal BP) and Ust Belai (7175 to 6320 cal BP) (*23*). In the Americas, the two Late Pleistocene specimens from Alaska, directly dated to 34,222 and 13,235 average cal BP, respectively (*24*), were classified by our analyses as morphological wolves. The earliest American morphological dog is from the Koster site in Illinois, directly dated to 8650 to 8250 cal BP (*25*) [Fig. 3A, (*3*)]. The status of the Koster specimen as a dog is supported by nuclear genomic evidence (*1*), and it is also one of the earliest confirmed dogs in the Americas and the earliest outside of Alaska (*26*).

**Hidden Late Pleistocene dogs**

To further refine the timing of when the morphological shift from wolves to dogs first appears, we conducted a comparative analysis within bounded time periods of contemporaneous canid specimens to the reference group of 86 modern wolves. Archaeological specimens were grouped in time bins of 50,000 to 25,000 cal BP, 25,000 to 10,000 cal BP, and then by consecutive 1000-year windows shifted by 100-year increments from 10,000 years BP to the present (Fig. 3, B and C). Because this approach is not based on our previously established identifications, it reveals periods when morphological selection was strong enough to deviate from the modern wild population.

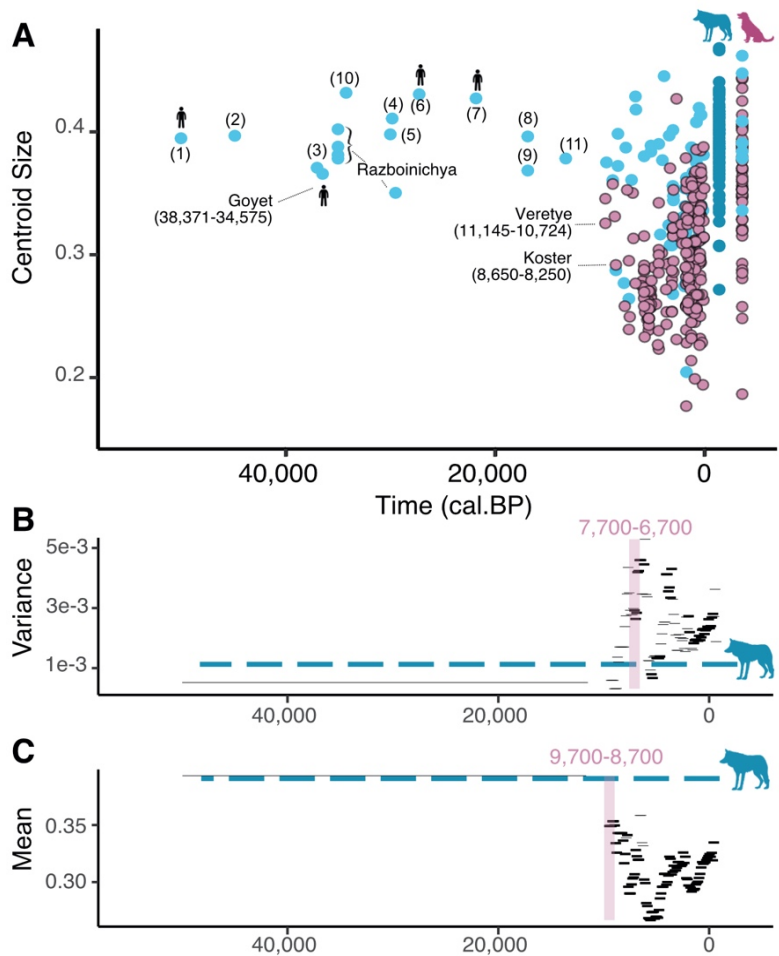

*Fig. 3. Diachronic morphometric changes.* (A) Variation in skull centroid size through time dividing the specimens between those identified as morphological dog (purple dots with black outline) and as morphological wolf (light blue). Modern wolf is shown in dark blue. The human profile indicates Pleistocene sites with confirmed human occupation. The sites are Grands Malades, Belgium (1); Banwell Cave, UK (2); Fanatic Cave, Russia (3); Kurtak, Russia (4); Badyarikkha, Russia (5); Maldidier, France (6); Trou des Nutons, Belgium (7); Eliseevichi, Russia (8); Ulakhan Sular, Russia (9); Fairbanks, Alaska (10); and Fairbanks, Alaska (11). (**B** and **C**) Sliding time windows comparing skull size variance (**B**) and mean (**C**). Bold black lines highlight significant differences, and the purple shaded area highlights the earliest time of differentiation. The blue dotted line represents the observed values for modern wolves.

Our results show that a reduction of skull size is first detectable from 9700 to 8700 years BP, an increase in size variance becomes noticeable from 7700 to 6700 years BP, and greater skull shape variability appears from 8200 to 7200 years BP (Fig. 3, B and C). In terms of mean skull shape, 97% of temporal groups of Holocene specimens were statistically different from modern wolves (all P < 0.05) (table S3). Combined, these results suggest that domestic dog morphology was present by the early Holocene and derived from existing wolf variability. None of the 17 Late Pleistocene specimens considered in this study, however, had a domestic morphology.

There are many reasons that the remains of Late Pleistocene dogs continue to be elusive. Carnivores are naturally rarer than herbivores in both living ecosystems and the archaeological record. Trophic constraints mean that fewer individuals can be supported at higher levels within the food chain. Additionally, because carnivores contribute little to human diets, their remains are less frequently deposited at (and thus recovered from) archaeological sites. When present, their skeletal elements, particularly fragile cranial bones, are often poorly preserved due to a range of taphonomic processes, including dog consumption. These biases affect both wolves and early dogs, making the assessment of morphological variability in Late Pleistocene canids especially challenging and likely underestimating their true diversity, which complicates the differentiation of wild and early domestic morphologies (*27*).

To assess the evolution of wolves through time, we examined both Late Pleistocene and modern wolf populations, which exhibit similar size but differ in their skull shape (correct cross-validated classification of 79.4%; CI: 70 to 86.8%, P = 0.001; fig. S2 and materials and methods). The skulls of modern wolves are also 26.4% less diverse than those of Late Pleistocene wolves (disparity tests: Z = 1.88, P = 0.032; table S2) (*28*). This greater diversity in Late Pleistocene wolves is expected because these specimens extend over a 30,000-year time period punctuated by major climatic shifts (*7*). The loss of morphological diversity and size reduction in recent wolves may also be linked to population decline, changes in habitat availability, and decreasing numbers of their large prey species throughout the Holocene (29). Human persecution of wolves across their native range, which was particularly severe during the Middle Ages, likely further reduced wolf diversity (30). In fact, our results demonstrate that modern specimens do not reflect the entire range of diversity observed in their ancient relatives. Skulls of morphological wolves from Holocene archaeological sites (*n* = 101) were on average smaller than those of modern wolves (fig. S4), with five specimens smaller than any of the Pleistocene or modern wolves included (table S1).

Among modern dog breeds, 20 specimens from breeds including the Malakli, German shepherd, St. Bernard, Tibetan mastiff, and some hound-like breeds exhibit a "wolflike" skull shape. Other individuals from these same breeds were categorized as "dog" (table S1). These results highlight the complexity of recognizing domestic individuals in the past. The ongoing search for Late Upper Pleistocene dogs is complicated by a combination of factors involving the recovery, preservation status, and analysis of canid remains from Pleistocene contexts. Especially critical is the analysis of specimens that predate Veretye, currently the earliest published genetically (7)—and here morphometrically—identified dog at 10,800 cal BP.

**Tracking domestic skull morphologies through time**

To characterize early dog skull morphologies, we compared the cranial size and shape of the 43 most ancient specimens identified as morpho- logical dogs from the Mesolithic and Neolithic to modern dogs (table S1). These archaeological specimens exhibit skull sizes and shapes that fall within the range observed in modern dogs. Overall, however, their skulls are smaller and less diverse in size on average than modern dogs (Fig. 4A), and the variability of their skull shape does not fully encom- pass the variation observed in the more numerous (n = 158) set of modern dog skulls (Fig. 4B). The modern dogs exhibit a wider range of extreme morphologies that includes short-faced (brachycephalic) bulldogs and pugs, as well as long-faced (dolichocephalic) borzois and Russian wolfhounds, which were not present in early archaeological specimens. Nevertheless, the diversity captured in the Mesolithic and Neolithic doubled (53%) that of Pleistocene specimens and represents half (52%) of the diversity present in modern dogs (table S4). This corroborates a recent study of 525 mandibles (8100 to 3000 cal BP) that also exhibited a significant proportion of modern variability (25).

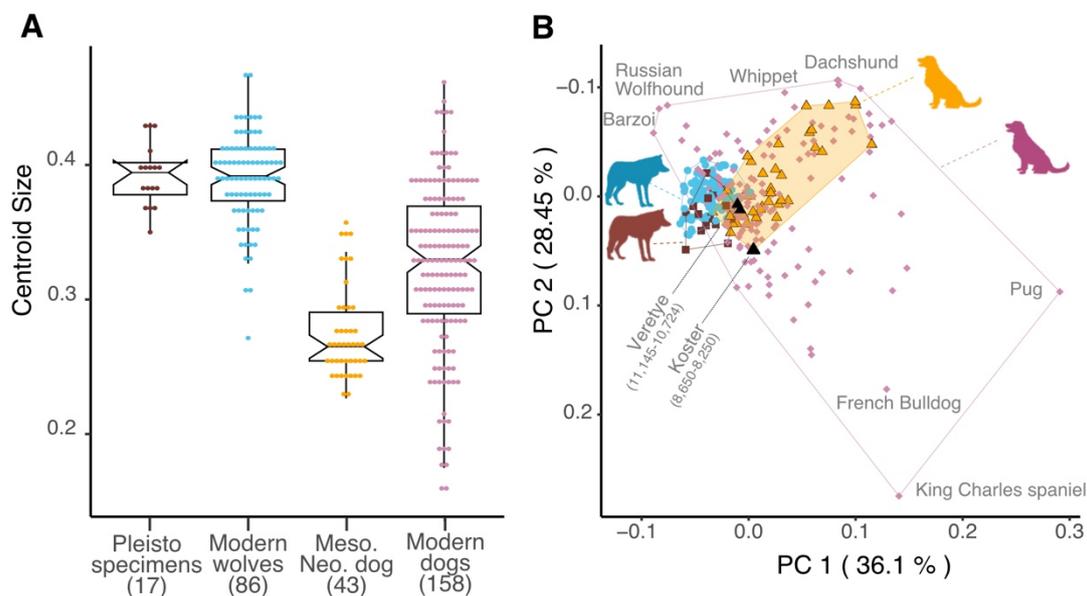

***Fig. 4. Morphometric variation of Mesolithic-Neolithic morphological dogs.*** *(A) Variation in skull centroid size. Mesolithic-Neolithic morphological dogs (n = 37) are smaller and less diverse in size than modern dogs (n = 158). (B) PCA on skull shape. Mesolithic-Neolithic morphological dogs (n = 37, orange triangles) had a more diverse skull shape than modern wolves (n = 86, blue circles) and Pleistocene wolves (n = 15, brown squares) but were less diverse than modern dogs (n = 158, purple diamonds). Breeds of the most extreme dog morphologies are depicted. Specimens from the Veretye (Russia) and Koster (Illinois) sites are depicted by solid black triangles.*

Together, these results demonstrate that whereas Victorian breeding programs are the origins of many of today's most extreme morphologies, early Holocene domestic dogs exhibited more diverse skull forms than previously considered corroborating previous studies based on an assessment of canid mandibles (*12*). This suggests that the pressures of human-induced selection alongside changing climatic conditions and food resource availability were sufficiently strong to lead to the appearance of diverse morphologies millennia before the definition of modern breeds.

**Conclusions**

The results presented here show that the earliest domestic dog skull morphology was present at 10,800 average cal BP at Veretye, the same Mesolithic Russian site where canids genomically identified as dogs have been identified (*21*). In addition, by the early Holocene, dogs exhibited a wide range of skull morphologies. Although all Pleistocene canids in our study had a skull morphology consistent with wolves, our results support a Pleistocene timing for wolf domestication. However, the earliest stages are not captured in our cranial dataset, and early dogs may have lived in close proximity with humans well before diversification of their skull morphology from wolves. The continued presence of wolflike dogs among modern breeds further highlights the complexity of disentangling the biological and cultural status of the earliest domes- tic individuals. Deepening our understanding of the origins of domestic dogs requires analyses of additional canid remains spanning the period from the Last Glacial Maximum (~26,000 to 20,000 years BP) to the end of the Pleistocene (11,700 years BP). The analysis of other skeletal elements, including mandibles, teeth, and postcranial bones, which are more numerous in faunal assemblages, would further expand our understanding of the morphological processes associated with domestication and adaptation to the anthropogenic niche, human selection, and other aspects of canid life histories (*31*).

Further analyses that focus on dog dispersal outside of the natural range of the wolf would clarify the biocultural history of this species during more recent periods. It would also support the investigation of lineages such as the dingo in Australia or the Xoloitzcuintle (hairless dog) in Central America, both of which have complex cultural and eco- logical relationships with humans. Additionally, combined biomolecular and archaeophenomic (*13*) studies could enhance our understanding of the natural and anthropic factors driving the emergence and diversification of dogs. The challenges faced in distinguishing domestic animals from their wild progenitors in the palaeontological and archaeological record reflect the continuum of domestication as a complex biocultural process shaped by intertwined biological, environmental, and cultural factors (*32*). Nowhere is this more true than for the study of the dog. Despite their long and diverse interaction with humans, dogs remain one of the most enigmatic and fascinating domestic species, not only as humanity's first domesticate but also as its most enduring companion.

**ACKNOWLEDGMENTS**
We gratefully acknowledge the many museums and institutions for providing access to their collections, the details of which can be found in table S1, as well as the individual contributions of
A. Hulme-Beaman, R.-M. Arbogast, A. Birley, P. Brewer, N. den Ouden, H.-J. Döhle, G. Font,
M. Germonpre, N. Hebdon, Y. Jing, J. Koler-Matznick, T. Lord, J. Mateu, R. Meadow,
K. Murphy Gregersen, M. Nowak Kemp, M. Nussbaumer, S. Pujadas, L. Peng, A. Perri, S. Peurach,
R. Portela Miguez, J. Tura, and B. Yates.
**Funding:** This work was supported by the Natural Environmental Research Council (grants NE/K005243/1, NE/K003259/1, NE/S007067/1, and NE/S00078X/1 to G.L.); the Arts and Humanities Research Council (UKRI grant AH/K006029/1 to R.K. and K.B.); the European Research Council (grant ERC-2019-STG-852573-DEMETER to A.E. and grant ERC-2013-StG-337574-UNDEAD to G.L.); the Social Sciences and Humanities Research Council of Canada (grant SSHRC IG435-2019-0706 to R.L.); the State Assignment of IGM SB RAS (grant 122041400252-1 to Y.K.); the State Assignment of ZIN RAS grant 125012800908-0 to M.S.); and the Fyssen Foundation (C.B.). **Author contributions:** Conceptualization: A.E., C.A., G.L., K.D.; **Data curation:** A.E., C.A., C.B. with input from all coauthors; Formal analysis: A.E.; **Funding acquisition:** G.L., K.D.; Methodology: A.E., C.A., J.C., V.B.; Visualization: A.E., C.B.; Writing – original draft: A.E., C.A.; Writing – review & editing: A.E., C.A., C.B., G.L., K.D. with input from all coauthors. **Competing interests:** The authors declare no competing interests. Data and materials availability: The generated 3D coordinates are provided in dataset S1. All other data are available in the supplementary materials.